\documentclass[%
  reprint,
 amsmath,amssymb,
 aps,
prb,
]{revtex4-1}

\usepackage[latin1]{inputenc}
\usepackage{hyperref}
\usepackage{natbib}
\usepackage{dcolumn}
\usepackage[dvipdfmx]{graphicx,psfrag}
\usepackage{color}
\usepackage{amsfonts,amsmath,bm}
\usepackage{braket}
\usepackage{subfigure}
\usepackage{ulem}

\newcommand{\FAL}{Fe$_3$Al}

\begin{document}

\title{
Enhancement of transverse thermoelectric conductivity originating from stationary points in nodal lines
}

\author{Susumu Minami}
\email{minami@cphys.s.kanazawa-u.ac.jp}
\affiliation{Nanomaterials Research Institute (NanoMaRi), Kanazawa University, Kakuma, Kanazawa, 920-1192, Japan}

\author{Fumiyuki Ishii}
\email{fishii@mail.kanazawa-u.ac.jp}
\affiliation{Nanomaterials Research Institute (NanoMaRi), Kanazawa University, Kakuma, Kanazawa, 920-1192, Japan}

\author{Motoaki Hirayama}
\affiliation{Center for Emergence Matter Science (CEMS), RIKEN, Hirosawa, Wako, Saitama 351-0198, Japan}

\author{Takuya Nomoto}
\affiliation{Department of Applied Physics, The University of Tokyo, Hongo, Bunkyo-ku, Tokyo, 113-8656, Japan}

\author{Takashi Koretsune}
\affiliation{Department of Physics, Tohoku University, Sendai, Miyagi, 980-8578, Japan}

\author{Ryotaro Arita}
\affiliation{Center for Emergence Matter Science (CEMS), RIKEN, Hirosawa, Wako, Saitama 351-0198, Japan}
\affiliation{Department of Applied Physics, The University of Tokyo, Hongo, Bunkyo-ku, Tokyo, 113-8656, Japan}

\date{\today}
\begin{abstract}
Motivated by the recent discovery of a large anomalous Nernst effect in Co$_2$MnGa, Fe$_3X$ ($X$=Al, Ga) and Co$_3$Sn$_2$S$_2$, we performed a first-principles study to clarify the origin of the enhancement of the transverse thermoelectric conductivity ($\alpha_{ij}$) in these ferromagnets. The intrinsic contribution to $\alpha_{ij}$ can be understood in terms of the Berry curvature ($\Omega$) around the Fermi level, and $\Omega$ is singularly large along nodal lines (which are gapless in the absence of the spin-orbit coupling) in the Brillouin zone. 
We find that not only the Weyl points but also stationary points in the energy dispersion of the nodal lines play a crucial role. The stationary points make sharp peaks in the  density of states projected onto the nodal line, clearly identifying the characteristic Fermi energies at which $\alpha_{ij}$ is most dramatically enhanced. We also find that $\alpha_{ij}/T$ breaks the Mott relation and show a peculiar temperature dependence at these energies. The present results suggest that the stationary points will give us a useful guiding principle to design magnets showing a large anomalous Nernst effect.
\end{abstract}

\maketitle
\section{Introduction}

In solids, a temperature gradient ($\bm{\nabla} T$) and an external electric field ($\bm{E}$) gives rise to a charge current ($\bm{J}$) given as 
\begin{equation}
{\bm J} = {\hat \sigma}{\bm E} + {\hat\alpha}(-\bm{\nabla} T),
\label{current}
\end{equation}
where $\hat\sigma$ and $\hat\alpha$ are the electric conductivity tensor and thermoelectric (TE) tensor, respectively. When ${\bm J}$ is absent, eq.~(\ref{current}) tells us that a temperature gradient generates a voltage as
\[
{\bm E}={\hat S}(\bm{\nabla} T),
\]
where ${\hat S}={\hat \sigma}^{-1} {\hat \alpha}$.
In ferromagnets, off diagonal elements of ${\hat\sigma}$ and ${\hat\alpha}$ are generally finite, so that a transverse voltage is induced by a longitudinal temperature gradient. This phenomenon is a thermoelectric counterpart of the anomalous Hall effect (AHE) and called the anomalous Nernst effect (ANE)~\cite{PhysRevLett.97.026603,RevModPhys.82.1539,RevModPhys.82.1959}.

Recently, the ANE is attracting renewed interest. It can be exploited in developing high-efficiency energy-harvesting devices with simple lateral structure, high flexibility and low production cost~\cite{SAKURABA201629,Mizuguchi_2019}.  
Experimental and theoretical studies of AHE~\cite{PhysRevLett.88.207208,Wang2018_NatComm,PhysRevLett.110.100401,Nayake1501870,Suzuki2016,Nakatsuji_Large_2015,PhysRevApplied.5.064009,PhysRevLett.112.017205,1367-2630-15-3-033014,PhysRevLett.92.037204,PhysRevX.8.041045,Li2020} and ANE~\cite{NatPhys_Co2MnGa,Guin2019,Fe3X_Nature,Liu2018_NatPhys,PhysRevMaterials.4.024202,doi:10.1002/adma.201806622,PhysRevB.98.205125,PhysRevB.96.224415,PhysRevApplied.13.054044,doi:10.1063/1.5029907,doi:10.1021/acs.nanolett.9b03739,PhysRevB.99.165117,doi:10.1063/1.4922901,PhysRevLett.93.226601,PhysRevLett.99.086602,PhysRevLett.101.117208,PhysRevLett.107.216604,Mn3Sn,FePt-MnGa,doi:10.1063/1.5143474,PhysRevMaterials.3.114412,PhysRevB.101.115106} have been reported in a variety of magnetic materials.
Among them, Co$_2$MnGa~\cite{NatPhys_Co2MnGa,Guin2019}, Fe$_3X$ ($X$=Al, Ga)~\cite{Fe3X_Nature} and
Co$_3$Sn$_2$S$_2$~\cite{Liu2018_NatPhys,PhysRevMaterials.4.024202,doi:10.1002/adma.201806622} are of particular interest due to their huge
anomalous transverse transport and less entangled low-energy electronic structure.

In fact, if the band dispersion around the Fermi level ($E_F$) is not so complicated, there is an intriguing possibility to design a giant ANE. This is because the transverse thermoelectric conductivity is directly related to the Berry curvature ($\bm\Omega$) of the low-energy bands, which can be calculated from first principles accurately~\cite{RevModPhys.82.1959}:
\begin{eqnarray}
  \sigma_{ij}(T,\mu)&=&-\varepsilon_{ijl}\frac{e^2}{\hbar}\int\frac{d\bm{k}}{(2\pi)^3}\sum_{m}\Omega_{m,l}(\bm{k})f(\varepsilon_{m\bm{k}}) \label{eq:sigma},\\
  \alpha_{ij}(T,\mu)&=&-\frac{1}{e}\int d\varepsilon \sigma_{ij}(0,\varepsilon)\frac{\varepsilon -\mu}{T} \left(-\frac{\partial f}{\partial \varepsilon}\right) \label{eq:alpha},
\end{eqnarray}
where $\varepsilon_{ijl}$, $e, \hbar, \varepsilon, f, \mu$ are the antisymmetric tensor, elementary charge with negative sign, the reduced Planck constant, the band energy, the Fermi-Dirac distribution function with the band index $m$ and the wave vector $\bm{k}$, and the chemical potential, respectively.
The Berry curvature for the $m$-th band is given as
\begin{equation}
{\Omega}_{m,l}(\boldsymbol{k}) = -2\varepsilon_{ijl}\operatorname{Im}{\sum_{m'\neq m}\frac{v_{mm',i}(\boldsymbol{k})v_{m'm,j}(\boldsymbol{k})  }{(\varepsilon_{m'}(\boldsymbol{k})- \varepsilon_m(\boldsymbol{k}) )^2}},
\label{Berry}
\end{equation}
where $v_{mm',i}$ denotes the matrix elements of the velocity operator along the $i$ direction, respectively.

In non-relativistic calculation, we generally find many nodal lines in the Brillouin zone.
The nodal line is a one-dimensional topological degeneracy where the energy gap closes~\cite{acs.nanolett.5b02978,PhysRevB.92.045108,PhysRevLett.115.036806,PhysRevLett.115.036807,1.4926545,PhysRevB.93.205132,JPSJ.85.013708,Hirayama2017,PhysRevB.93.201114}.
The spin-orbit coupling (SOC) opens a small gap along these nodal lines and the Berry curvature is singularly large there because the energy difference $|\varepsilon_{m'}-\varepsilon_m|$ is small (see
eq.~(\ref{Berry})). In particular, the Berry curvature diverges at the Weyl points, at which the band crossing survives even in the presence of the SOC.
Therefore, the existence of the nodal lines and especially the Weyl points around $E_F$ has been considered to be critically important for realizing large anomalous transverse transport. Indeed, the role of various topological objects such as the type-II Weyl point~\cite{soluyanov2015_nature} and Hopf link of nodal lines have been extensively studied for Heusler ferromagnets~\cite{NatPhys_Co2MnGa,Co2MnGa_Hasan_PRL,Guin2019,Fe3X_Nature,PhysRevB.99.165117,PhysRevX.8.041045,Manna2018,Li2020},
Co$_3$Sn$_2$S$_2$~\cite{Liu2018_NatPhys,PhysRevMaterials.4.024202,doi:10.1002/adma.201806622}, and other ferromagnets~\cite{Kim_2018,PhysRevB.98.245132,Kanazawa_2016,Destraz2020}.
However, there is no established general guiding principle to design magnets showing a large ANE.

In this paper, we propose that stationary points in the energy dispersion of nodal lines play a crucial role to determine the best energy for $\mu$ at which the ANE is most dramatically enhanced. The nodal lines are one-dimensional objects in the Brillouin zone, so that the stationary points make sharp peaks in the density of states (DOS) projected onto the nodal lines:
\[
D_{\rm NL}(\varepsilon)=\sum_{n,{\bm k}\in{\bm k}_{\rm NL}} \delta(\varepsilon-\varepsilon_{n{\bm k}}),
\]
where $n$ is the band index and ${\bm k}_{\rm NL}$ specifies the positions of the nodal lines.
Based on first-principles calculations for Co$_2$MnGa, Fe$_3$Al and Co$_3$Sn$_2$S$_2$, we show that there is a clear one-to-one correspondence between the ``van Hove singularities'' in the $D_{\rm NL} (\varepsilon=E_{\rm VHS}$) and the energy for $\mu$ at which the transverse TE conductivity is enhanced.

When $\mu$ is located at $E_{\rm VHS}$, we can also find a breakdown of the Mott relation as a prominent indication of the enhancement of the transverse TE conductivity. The Mott relation is derived by using the Sommerfeld expansion for eq.~(\ref{eq:alpha}), which is usually valid at sufficient low temperatures~\cite{ziman_1972,mott1958theory,PhysRevB.21.4223,PhysRev.181.1336,PhysRevB.64.224519,PhysRevLett.101.117208,PhysRevLett.100.106601,PhysRevLett.99.086602}: 
\[
\alpha_{ij}(T,\mu) = \left.-\frac{\pi^2k_{\rm B}^2T}{3|e|}\frac{d{\sigma_{ij}}(0,\varepsilon)}{d\varepsilon}\right|_{\varepsilon=E_{\rm F}}, \
\]
where $k_{\rm B}$ is the Boltzmann constant.
Thus for many materials, $\alpha_{ij}/T$ is a constant at $T\rightarrow 0$. However, recently, it has been found that $\alpha_{ij}/T$ diverges at low temperatures in several ferromagnets which exhibit a large ANE~\cite{NatPhys_Co2MnGa,PhysRevMaterials.4.024202,Liu2018_NatPhys,PhysRevB.98.205125,PhysRevLett.97.026603,PhysRevB.96.224415,Fe3X_Nature}. We show that this peculiar behavior can be understood in terms of $D_{\rm NL}$:
The energy dependence of ${\sigma_{ij}}(0,\varepsilon)$ is singular
at $\varepsilon=E_{\rm VHS}$ where $D_{\rm NL}$ has a sharp peak. There, the Sommerfeld expansion does not work even at low temperatures.
We show that the Mott relation is indeed violated for Co$_2$MnGa, Fe$_3$Al and Co$_3$Sn$_2$S$_2$, when $\mu$ is close to $E_{\rm VHS}$ and the transverse TE conductivity is strongly enhanced.

\section{Computational details} \label{sec:method}

\begin{table}[tb] \centering
  \caption{\label{tab:mate_info} Space group, lattice constant ($a,c$), and Curie temperature ($T_{\rm C}$) of each material. Our calculations were performed using the experimental lattice constants.}
  \begin{tabular}{cccc} \hline \hline
    \it{M} & space group  & $a,c$ (\AA) & $T_{\rm C}$ (K) \\ \hline 
    Co$_2$MnGa        & $Fm\bar{3}m$ & 5.77 \footnote{Ref.~[\onlinecite{NatPhys_Co2MnGa}]}            & 694    \footnote{Ref.~[\onlinecite{WEBSTER19711221}]}     \\ 
    Co$_3$Sn$_2$S$_2$ & $R\bar{3}m $ & $a=5.36,c=13.17$ \footnote{Ref.~[\onlinecite{Liu2018_NatPhys}]}   & 177   \footnote{Ref.~[\onlinecite{VAQUEIRO2009513}]}      \\
    Fe$_3$Al          & $Fm\bar{3}m$ & 5.79 $^{\rm e}$     & 760    \footnote{Ref.~[\onlinecite{PhysRevLett.79.1909}]} \\ \hline \hline
  \end{tabular} 
  \end{table}

We conducted first-principles calculations based on the non-collinear density functional theory \cite{NC1} (DFT) with {\sc OpenMX} code \cite{OpenMX}.
DFT calculations are performed through the exchange-correlation functional within the generalized-gradient approximation and norm-conserving pseudopotentials \cite{PhysRevB.47.6728}. 
The SOC is included by using total angular momentum dependent pseudopotentials  \cite{PhysRevB.64.073106}.
The wave functions are expanded by a linear combination of multiple pseudo-atomic orbitals \cite{PhysRevB.67.155108}.
A set of pseudoatomic orbital basis was specified as Al7.0-$s3p3d1$, S7.0-$s3p3d1$, Mn6.0-$s3p3d3$, Fe6.0-$s3p3d3$, Co6.0-$s3p3d3$, Ga7.0-$s3p3d3$, and Sn7.0-$s3p3d1$ where the number after each element stands for the radial cutoff in the unit of bohr and the integer after $s, p, d$ indicates the radial multiplicity of each angular momentum component.
The cutoff energy for charge density of 800 Ry and  a $k$-point mesh of $35\times35\times35$ were used.
The nodal lines were obtained by monitoring the degeneracy of eigenvalues in the momentum space based on electronic structure without SOC~\cite{WU2017}.
Table \ref{tab:mate_info} shows the space group, lattice constant, and Curie temperature of each material.
The lattice constants of each material refer to the experimental ones  as listed in Table  \ref{tab:mate_info}.

From the Bloch states obtained in the DFT calculation, a Wannier basis set was constructed by using the {\sc Wannier90} code\cite{Pizzi2020}.
The basis was composed of ($s,p$)-character orbitals localized at the Al and S site, $d$-character orbitals at the Co, Mn and Fe site, $p$-character orbitals at the Ga and Sn site. Therefore, we consider 36 orbitals/f.u. for Co$_2$MnGa, 62 orbitals/f.u. for Co$_3$Sn$_2$S$_2$, 38 orbitals/f.u. for Fe$_3$Al including the spin multiplicity.
These sets were extracted from 194, 102 and 92 bands 
in the energy window ranging from $-20$ eV to $+50$ eV, $-15$ eV to $+40$ eV, and $-15$ eV to $+50$ eV for Co$_3$Sn$_2$S$_2$, Co$_2$MnGa, and Fe$_3$Al, respectively.

The anomalous Hall conductivity (eq.~(\ref{eq:sigma})) and the anomalous transverse TE conductivity (eq.~(\ref{eq:alpha})) at finite temperature were computed with the {\sc Wannier90} code using a $k$-point mesh of $100\times100\times100$ and additionally an adaptive mesh of $3\times3\times3$ for regions with large $\Omega_{n,l}$.

\section{Results and discussion}

\subsection{Enhancement of transverse thermoelectric conductivity and violation of the Mott relation}

\begin{figure}[tbp] \centering
  \includegraphics[width=\columnwidth]{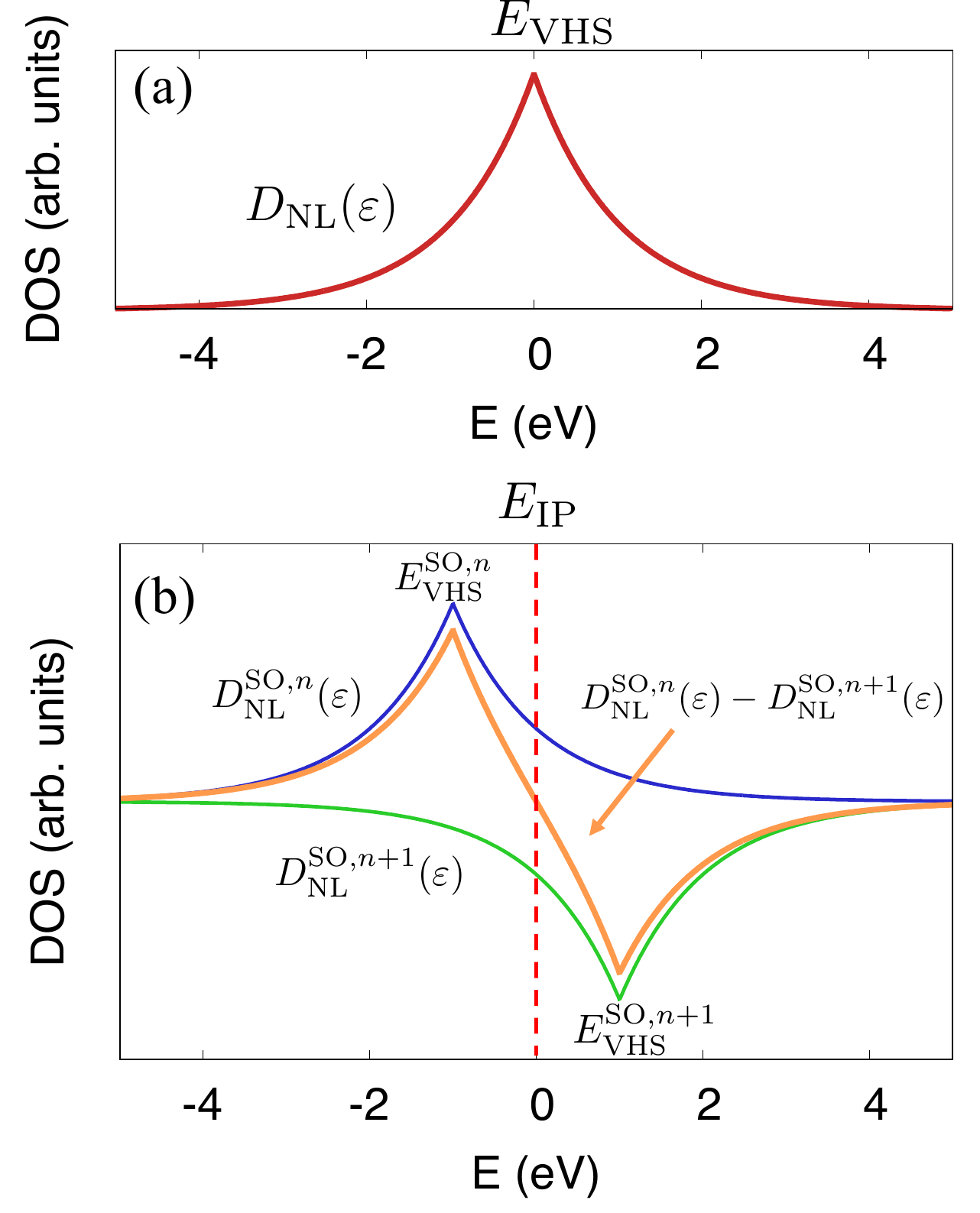}   
  \caption{
  Schematic of (a) $D_{\rm NL}(\varepsilon)$ without SOC and (b) $D_{\rm NL}^{\rm SO, n}$, $-D_{\rm NL}^{\rm SO, n+1}$ with SOC
  and $D_{\rm NL}^{\rm SO, n}-D_{\rm NL}^{\rm SO, n+1}$ in eq~(\ref{dosdiff}).
  }
  \label{fig:DOSNL} 
\end{figure}

Let us first discuss how the van Hove singularities in $D_{\rm NL}$ cause the enhancement of the transverse TE conductivity and breakdown of the Mott relation.
Equation~(\ref{eq:alpha})
can be rewritten as~\cite{PhysRevLett.97.026603} 
\begin{eqnarray}
  \alpha_{ij} &=& \frac{k_{\rm B}}{e} \int d \varepsilon \frac{\partial \sigma_{ij}}{\partial \varepsilon}  s\left(\varepsilon,T \right),
  \label{eq:alpha2}\\
\frac{\partial \sigma_{ij}}{\partial \varepsilon} 
&=&\varepsilon_{ijl}\sum_{n\bm{k}}\Omega_{n,l}({\bm k})\delta(\varepsilon-\varepsilon_{n\bm{k}}),
\label{eq:derivativeepsilon}
\\
s  &=& -f\ln(f) - (1-f)\ln(1-f),
\end{eqnarray}
where $s(\varepsilon,T)$ is the entropy density.

For eq.~(\ref{eq:derivativeepsilon}), let us introduce the following approximation: In the presence of the SOC, the absolute value of the Berry curvature $\Omega_{n,l}({\bm k})$ is large at ${\bm k}$ where the effect of the SOC on $\varepsilon_{n\bm{k}}$ is prominent. Thus $|\Omega_{n,l}({\bm k})|$ takes a large value on nodal lines in the non-relativistic calculation for which the SOC opens a gap.
Suppose that the $n$-th and $n+1$-th band make nodal lines (at ${\bm k}\in{\bm k}_{\rm NL}$) in the absence of the SOC and consider the density of states $D_{\rm NL}(\varepsilon)$
for these band crossing points.
The SOC splits $D_{\rm NL}(\varepsilon)$ into the DOS for the $n$-th band ($D^{{\rm SO},n}_{\rm NL}(\varepsilon)$) and $n+1$-th band ($D^{{\rm SO},n+1}_{\rm NL}(\varepsilon)$). Since $|\Omega_{n,l}({\bm k})|$ is small at ordinary ${\bm k}$ points that are not on the nodal lines (i.e., ${\bm k}\notin{\bm k}_{\rm NL}$),

\begin{eqnarray}
    \frac{\partial \sigma_{ij}}{\partial \varepsilon} \sim \Omega_{\rm NL}^{n}(\varepsilon)D^{{\rm SO},n}_{\rm NL}(\varepsilon)+\Omega_{\rm NL}^{n+1}(\varepsilon)D_{\rm NL}^{{\rm SO},n+1}(\varepsilon). 
    \label{eq:approx1}
\end{eqnarray}
Here, $\Omega^n_{\rm NL}(\varepsilon)$ denotes the averaged value of the Berry curvature on the nodal line,
\begin{equation}
\Omega_{\rm NL}^{n}(\varepsilon) = \sum_{{\bm k}\in{\bm k}_{\rm NL}}\Omega_{n,l}({\bm k}) \delta(\varepsilon - \varepsilon_{n\boldsymbol{k}}) / \sum_{{\bm k}\in{\bm k}_{\rm NL}} \delta(\varepsilon - \varepsilon_{n\boldsymbol{k}}). \label{averageOmega}
\end{equation}
When $\boldsymbol{k}$ is on the nodal line formed by the $n$-th and $n+1$-th band, the contribution of $m=n$, $m'=n+1$ is prevailing in Eq.~\eqref{Berry} for $\Omega_{n,l}(\boldsymbol{k})$ since the factor of $1/(\varepsilon_{n \boldsymbol{k}} - \varepsilon_{n+1 \boldsymbol{k}})^2$ is dominantly large. Similarly for $\Omega_{n+1,l}(\boldsymbol{k})$, the contribution of $m=n+1$, $m'=n$ is dominant.
Thus, if we assume $\varepsilon_{n\boldsymbol{k}} \sim \varepsilon_{n+1\boldsymbol{k}}$, then $\Omega_{\rm NL}^{n}(\varepsilon) \sim -\Omega_{\rm NL}^{n+1}(\varepsilon)$, and eq.~(\ref{eq:approx1}) can be further approximated as
\begin{eqnarray}
\frac{\partial \sigma_{ij}}{\partial \varepsilon}
    \sim \Omega^n_{\rm NL}(\varepsilon)(D^{{\rm SO},n}_{\rm NL}(\varepsilon)-D_{\rm NL}^{{\rm SO},n+1}(\varepsilon)) \label{dosdiff}.
\end{eqnarray}

We illustrate a schematic of $D_{\rm NL}$ with and without SOC in Fig.~\ref{fig:DOSNL}. 
Since $D_{\rm NL}$ is essentially the DOS of one-dimensional objects, it has sharp peaks (``van Hove singularities'') at the energies of stationary points (i.e, $\varepsilon=E_{\rm VHS}$) in the nodal lines as shown in Fig.~\ref{fig:DOSNL}(a).
In the presence of the SOC, $D_{\rm NL}^{{\rm SO},n}$ and $D_{\rm NL}^{{\rm SO},n+1}$ also have sharp peaks at $\varepsilon=E_{\rm VHS}^{{\rm SO},n}, E_{\rm VHS}^{{\rm SO},n+1}$ as shown in Fig.~\ref{fig:DOSNL}(b). 
Since $s_{n{\bm k}}$ takes a maximum around $\varepsilon=E_{\rm F}$, we see that $\alpha_{ij}$ will be enhanced when $E_{\rm VHS}^{\rm SO,n}=E_{\rm F}$ (see eq.~(\ref{eq:alpha2})).

As we have seen in eq.~(\ref{dosdiff}), $\partial \sigma_{ij}/\partial \varepsilon$ is approximately proportional to the difference between $D_{\rm NL}^{{\rm SO},n}(\varepsilon)$ and $D_{\rm NL}^{{\rm SO},n+1}(\varepsilon)$ (See Fig.~\ref{fig:DOSNL}(b)). Thus we expect that $\alpha_{ij}$ takes its maximum or minimum at  $E_{\rm VHS}^{{\rm SO},n}$ $(E_{\rm VHS}^{{\rm SO},n+1})$  in $D_{\rm NL}^{{\rm SO},n}(\varepsilon)$ ($D_{\rm NL}^{{\rm SO},n+1}(\varepsilon)$). On the other hand, $E_{\rm VHS}$ in $D_{\rm NL}(\varepsilon)$ is located between those in $D_{\rm NL}^{{\rm SO},n}(\varepsilon)$ and $D_{\rm NL}^{{\rm SO},n+1}(\varepsilon)$. Therefore, $E_{\rm VHS}$ in $D_{\rm NL}(\varepsilon)$ is expected to reside between the minimum and maximum in $\alpha_{ij}$. Namely, $E_{\rm VHS}$ in $D_{\rm NL}(\varepsilon)$ corresponds to the ``inflection point'' in $\alpha_{ij}$ and gives crucial information to identify the chemical potential at which $\alpha_{ij}$ is substantially enhanced.

We can further show that if $D^{{\rm SO},n}_{\rm NL}$ or $D^{{\rm SO},n+1}_{\rm NL}$ has a logarithmic singularity at $\varepsilon=E_{\rm VHS}$ and $E_{\rm VHS}=E_{\rm F}$, $\alpha_{ij}/T=c_1 \ln T +c_2$.
Similarly, if $D^{\rm SO,n}_{\rm NL}(\varepsilon)$ or $D^{\rm SO,n+1}_{\rm NL}(\varepsilon)$ is proportional to $(\varepsilon-E_{\rm VHS})^m$, $\alpha/T=c_3 T^m$.
Here, $c_1$, $c_2$ and $c_3$ are constants which do not depend on $T$. In Ref.~[\onlinecite{NatPhys_Co2MnGa}], it has been proposed that when the Weyl fermions reside close to the Lifshitz transition from the type-I to type-II, 
$\partial \sigma_{ij}/\partial \varepsilon$ has a logarithmic divergence, which leads a quantum critical behavior of the transverse TE conductivity.
Our present discussion is a generalization of this result.

\subsection{Magnetic Weyl semimetal Co$_3$Sn$_2$S$_2$}

\begin{figure}[tbp] \centering
  \includegraphics[width=\columnwidth]{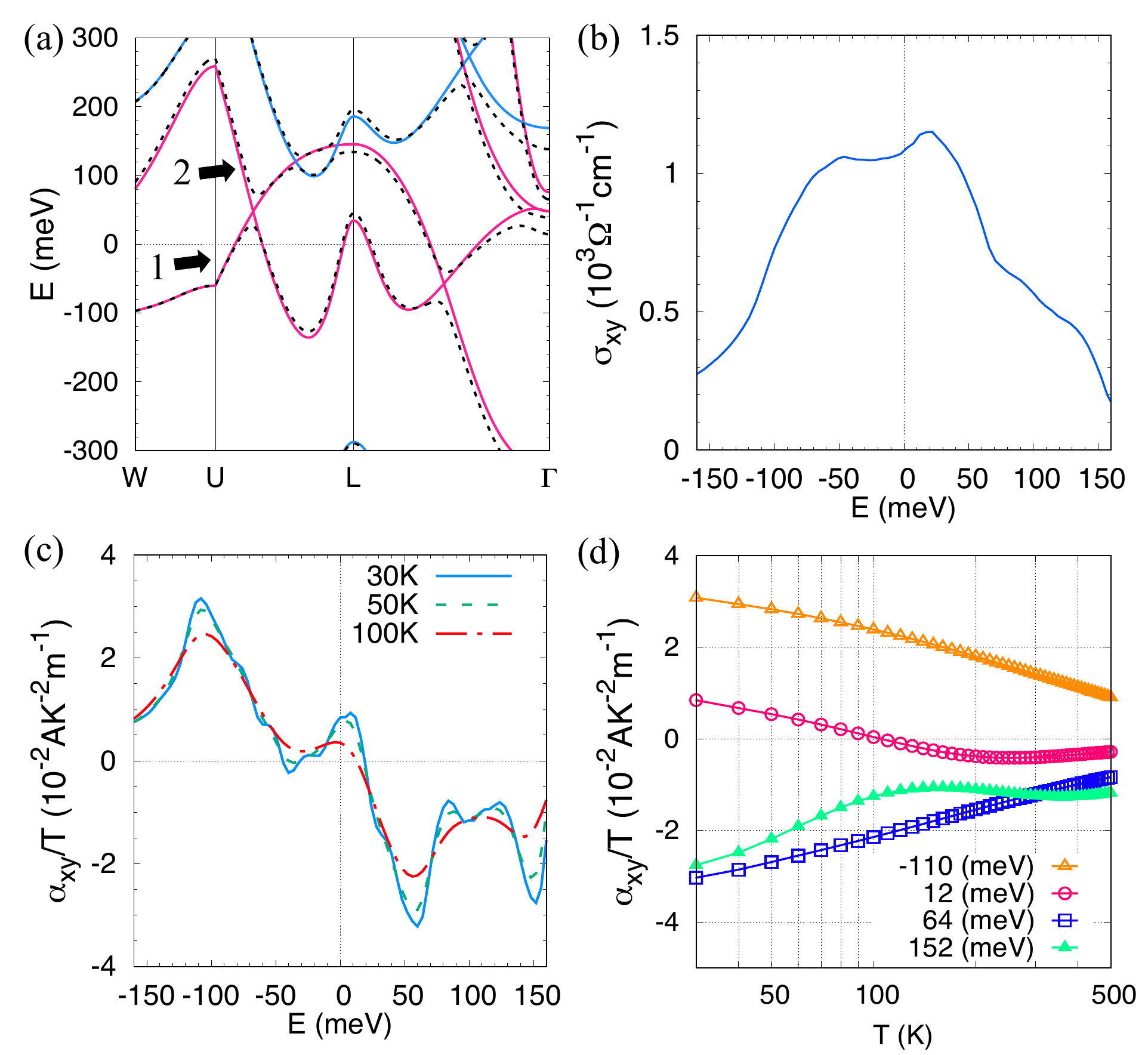}   
  \caption{(a) Band structure of Co$_3$Sn$_2$S$_2$. Pink and cyan lines correspond to the majority and minority spin bands computed without SOC. Dashed lines show the band calculated including SOC. 
  The denoted majority bands 1, and 2 make the nodal line.
  (b) $\mu$ dependence of $\sigma_{xy}$ at 0~K. (c) $\mu$ dependence of $\alpha_{xy}/T$. 
  Solid, dotted, and dash-dotted line are the results for $T=$ 30, 50, and 100K, respectively.
  (d) $T$ dependence of $\alpha_{xy}/T$.
  Open triangle, open circle, open square and solid triangle line correspond to the results for $\mu=-110$, $12$, $64$, and $152$ meV, respectively.}
  \label{fig:Co3Sn2S2-1} 
\end{figure}

\begin{figure}[tbp] \centering
  \includegraphics[width=\columnwidth]{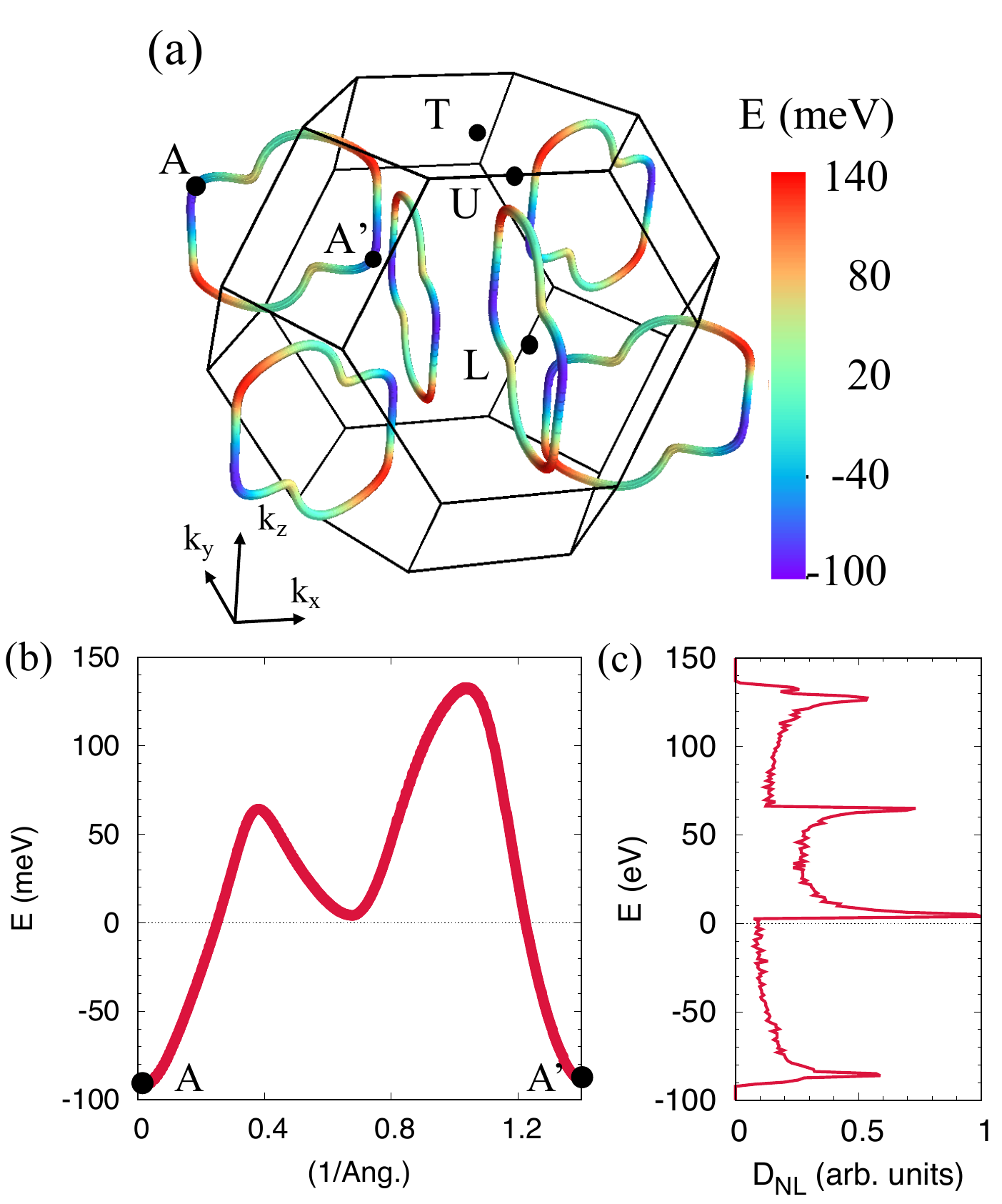}
  \caption{(a) Nodal line network of Co$_3$Sn$_2$S$_2$. The color bar corresponds to the energy range from $-100$ to $140$ meV.
  (b) Energy dispersion along the nodal line and (c) $D_{\rm NL}$ for Co$_3$Sn$_2$S$_2$. The A and A' point are shown in Fig. \ref{fig:Co3Sn2S2-2}(a).}
  
  \label{fig:Co3Sn2S2-2} 
\end{figure}

\begin{figure}[tbp] \centering
  \includegraphics[width=\columnwidth]{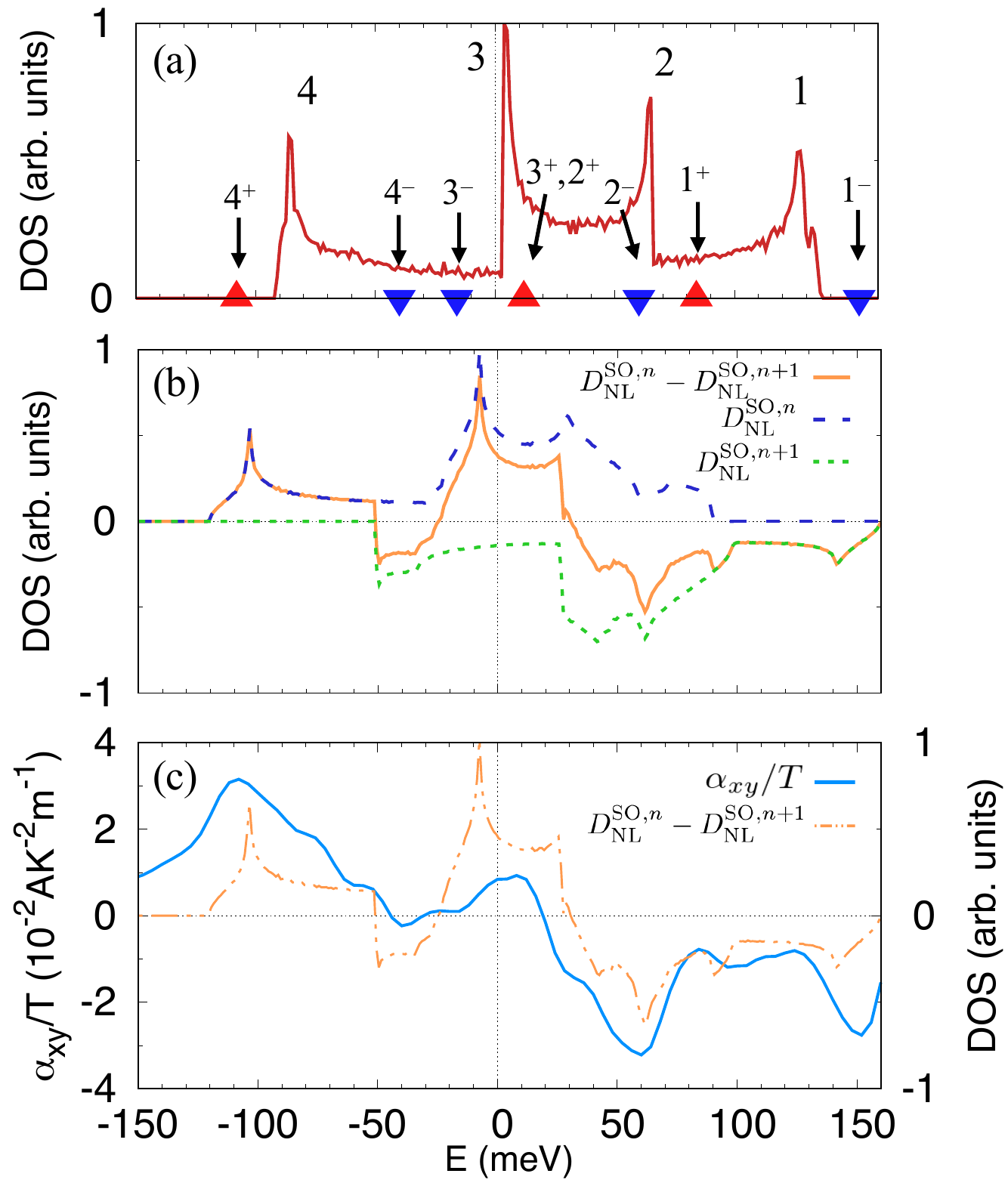}
  \caption{  (a) $\mu$ dependence of $D_{\rm NL}$. The peaks 1, 2, 3, and 4 denote $E_{\rm VHS}$.
  Solid upper (lower) triangle shows the maximum (minimum) in $\alpha_{xy}/T$. Each value of energies is tabulated in Tab.~\ref{tab:nldos_co3sn2s2-2}.
  (b) $\mu$ dependence of $D_{\rm NL}$ with SOC. Solid, dashed, dotted line are  $D_{\rm NL}^{{\rm{SO},n}}-D_{\rm NL}^{{\rm SO},n+1}$, $D_{\rm NL}^{{\rm SO},n}$, and $-D_{\rm NL}^{{\rm SO},n+1}$ in eq.~(\ref{dosdiff}), respectively. 
  (c) $\mu$ dependence of $\alpha_{xy}/T$ and $D_{\rm NL}^{{\rm{SO},n}}-D_{\rm NL}^{{\rm SO},n+1}$. Solid line is the results for $T=30$ K. Dashed double-dotted line shows $D_{\rm NL}^{{\rm{SO},n}}-D_{\rm NL}^{{\rm SO},n+1}$.
  }
  \label{fig:Co3Sn2S2-3} 
\end{figure}

Let us now look into the first-principles calculation for the magnetic Weyl semimetal Co$_3$Sn$_2$S$_2$, which has simple electronic structure composed of Weyl points and nodal line near the $E_{\rm F}$.
Figure \ref{fig:Co3Sn2S2-1} shows the band structure, $\mu$ dependence of $\sigma_{xy}$ and $\alpha_{xy}/T$, and $T$ dependence of $\alpha_{xy}/T$.
The obtained total magnetic moment is $ 0.9$ $\mu_{\rm B}$/f.u. so that spin polarization per each Co atom is $ 0.3$ $\mu_{\rm B}$.
The calculated $\sigma_{xy}$ at $E_{\rm F}$ is 1093 $\Omega^{-1}$cm$^{-1}$, which is consistent with the previous theoretical calculation (1100 $\Omega^{-1}$cm$^{-1}$) and the experimental value (1130 $\Omega^{-1}$cm$^{-1}$)~[\onlinecite{Liu2018_NatPhys}].
Figures \ref{fig:Co3Sn2S2-1}(c) and (d) show that $\alpha_{xy}/T$ is enhanced and has a peculiar temperature dependence at $\mu=-110$, 12, 64 and 152 meV.
Namely, $\alpha_{xy}/T$ does not satisfy the Mott relation at these energies.

As is seen in Fig. \ref{fig:Co3Sn2S2-1}(a), 
the low-energy electronic structure of Co$_3$Sn$_2$S$_2$ is very simple:
Only the majority spin contributes to the Fermi surface and 
there is only one nodal line formed by band 1 and band 2 in Fig.~\ref{fig:Co3Sn2S2-1}(a).

Let us next discuss the enhancement of $\alpha_{ij}$ and the violation of the Mott relation in terms of the nodal line.
In Fig.~\ref{fig:Co3Sn2S2-2}(a), we show the nodal line in the Brillouin zone.
We can see that the ``band width'' of the nodal line is about 240 meV, and the nodal line appears on the high symmetry planes.
Figures \ref{fig:Co3Sn2S2-2}(b) and \ref{fig:Co3Sn2S2-2}(c) show the energy dispersion along the nodal line (the ``nodal-line band'') and $D_{\rm NL}$ for Co$_3$Sn$_2$S$_2$, respectively. We chose the $k$-path along the nodal line in Fig.~\ref{fig:Co3Sn2S2-2}(a), where the positions of the A and A' are indicated.
Due to the symmetry of the Brillouin zone, there are two periods of changes in the Berry curvature in one loop of the nodal line. The nodal-line band in Fig. \ref{fig:Co3Sn2S2-2}(b) have one maximum, one minimum and two other stationary points.
The energies of these points have a one-to-one correspondence with the van Hove singularities in $D_{\rm NL}$ (Fig. \ref{fig:Co3Sn2S2-2}(b)).

Figure~\ref{fig:Co3Sn2S2-3} shows the $\mu$ dependence of  $D_{\rm NL}$ and $D_{\rm NL}^{{\rm SO},n}$, $D_{\rm NL}^{{\rm SO},n+1}$ and $D_{\rm NL}^{{\rm{SO},n}}$ $-D_{\rm NL}^{{\rm{SO},n+1}}$, and  $\alpha_{xy}/T$.
In Fig.~\ref{fig:Co3Sn2S2-3}(a), we see that four sharp peaks in $D_{\rm NL}$.
In Fig.~\ref{fig:Co3Sn2S2-3}(b), we plot $D_{\rm NL}^{{\rm{SO},n}}$, $-D_{\rm NL}^{{\rm{SO},n+1}}$, and $D_{\rm NL}^{{\rm{SO},n}}-D_{\rm NL}^{{\rm{SO},n+1}}$ and compare the energies of the van Hove singularities ($E_{\rm VHS}$'s). 
In Fig.~\ref{fig:Co3Sn2S2-3}(c), we see that the peaks in $D_{\rm NL}^{{\rm SO},n}$ and $-D_{\rm NL}^{{\rm SO},n+1}$ correspond to the energies at which $\alpha_{xy}/T$ takes its maximum and minimum, respectively.
These peaks originate from the ``van Hove singularites'', i.e., the stationary points in the energy dispersion of the nodal lines.
We see that each $E_{\rm VHS}$ in $D_{\rm NL}$ is located between those in $D_{\rm NL}^{{\rm SO},n+1}$ and $D_{\rm NL}^{{\rm SO},n}$.
Therefore, each $E_{\rm VHS}$ in $D_{\rm NL}$ corresponds to the "inflection point" between the maximum and minimum in $\alpha_{xy}/T$.

In Table \ref{tab:nldos_co3sn2s2-2}, we compare the inflection point in $\alpha_{xy}/T$ and $E_{\rm VHS}$ in $D_{\rm NL}$ more explicitly. The energy of the inflection point $E_{\rm IP}$ is estimated by taking the average of the energies at which $\alpha_{xy}/T$ takes its maximum and minimum and shows the breakdown of the Mott relation. 
While there is some deviation $\sim$ 10 meV between these two characteristic energies, we see that the peak 1, 3, and 4 clearly correspond to $E_{\rm IP}$. As for the origin of the deviation, we should note that $E_{\rm VHS}$ in $D_{\rm NL}$ is determined by a calculation without SOC.
While the correspondence is not so clear for peak 2 (which is due
to the presence of the Weyl points at $\sim$ 60 meV\cite{Liu2018_NatPhys}),
we can conclude that a divergence in $D_{\rm NL}$ enhances $\alpha_{xy}/T$ and causes the breakdown of the Mott relation.

While we have seen that $\alpha_{xy}/T$ is always enhanced when $\mu$ is located around peaks of $D_{\rm NL}^{{\rm SO},n+1} (D_{\rm NL}^{{\rm SO},n})$, it is difficult to predict the absolute value of $\alpha_{xy}/T$ by just looking at the value of $D_{\rm NL}$. For example, although peak 3 in $D_{\rm NL}$ is higher than peak 2 (Fig.~\ref{fig:Co3Sn2S2-2}(b)), the absolute value of $\alpha_{xy}/T$ at 64 meV (which corresponds to peak 2) is larger than that at 12 meV (which corresponds to peak 3). This is because the averaged $\Omega^n_{\rm NL}$ (eq.~(\ref{averageOmega})) in peak 2 is larger than that in peak 3. As is mentioned above, it has been shown that there are Weyl points at $\sim$60 meV, which generally make $\Omega^n_{\rm NL}$ larger~\cite{Liu2018_NatPhys}.

\begin{table}[tbp] \centering
 \caption{One-to-one correspondence between the peaks in $D_{\rm NL}$ and $E_{\rm IP}$ in Co$_3$Sn$_2$S$_2$. 
 $E_{\rm IP}$ is estimated as an average of energy taking the maximum and minimum in  $\alpha_{xy}/T$. 
 $\alpha_{xy}^{+(-)}/T$ denotes the energy at which $\alpha_{xy}/T$ takes its maximum (minimum) and deviates from the Mott relation. }
 \begin{tabular}{ccccc} \hline \hline
 Peak  & $D_{\rm NL}$ (meV) & $E_{\rm IP}$ (meV)   & $\alpha_{xy}^{+}/T$ (meV) &  $\alpha_{xy}^{-}/T$ (meV) \\ \hline 
    1     &  126               &  118                   & 84 & 152                  \\
    2     &   62               &  36                   & 12 &  60                  \\
    3     &    3               &   -2                   &   12 & -16                  \\
    4     &  -88               &  -74                   & -108 & -40                  \\ \hline \hline
    \end{tabular} 
  \label{tab:nldos_co3sn2s2-2}
  \end{table}

\subsection{Magnetic Weyl semimetal Co$_2$MnGa}

\begin{figure}[tbp] \centering
  \includegraphics[width=\columnwidth]{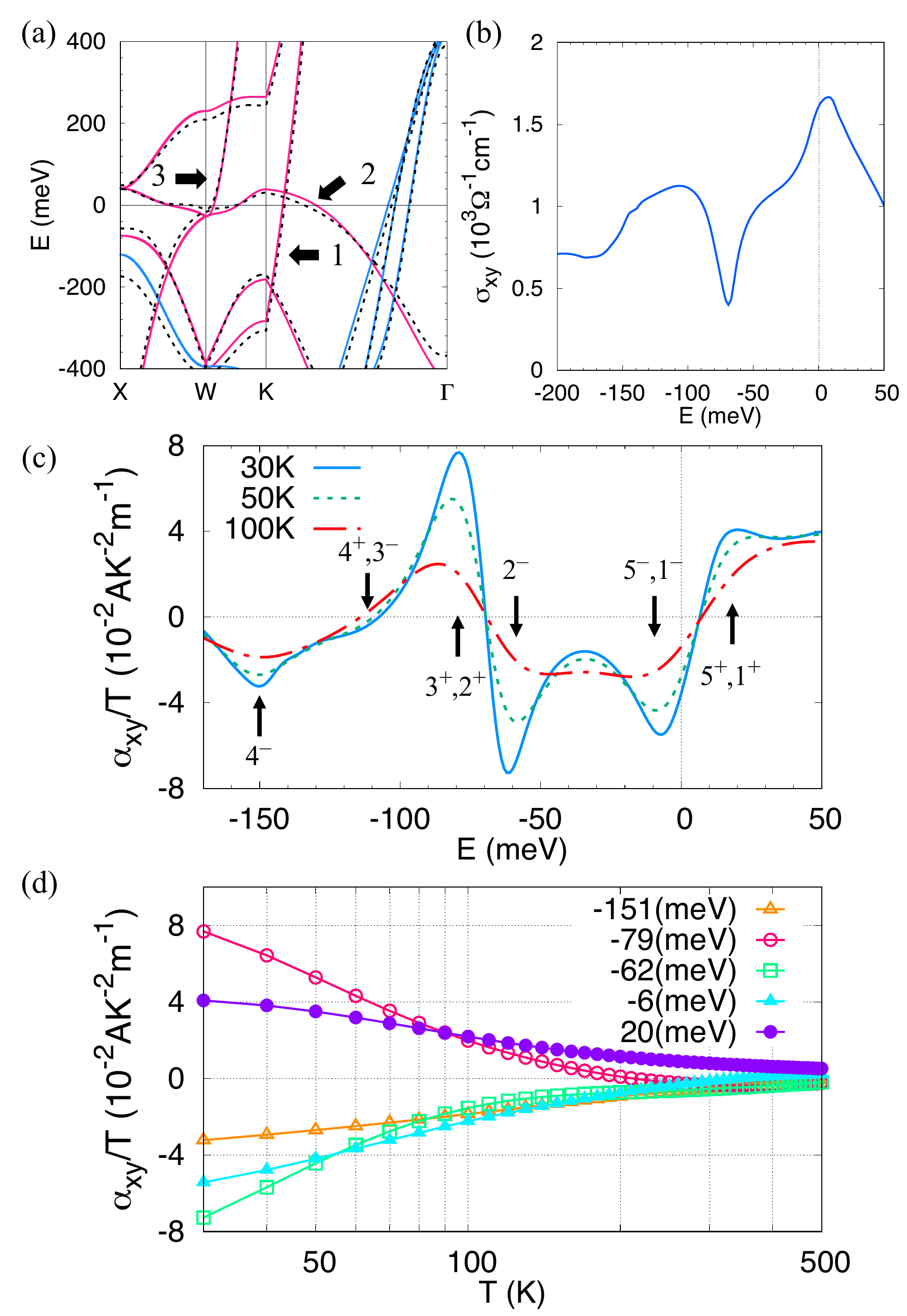}   
  \caption{(a) Band structure of Co$_2$MnGa. Pink and cyan lines correspond to up and down spin bands computed without SOC. Dashed lines show the band calculated including SOC. $E_F$ and $\mu$ for the doped case are measured from the original Fermi level. The denoted majority bands 1, 2, and 3 make the nodal lines.
  (b) Chemical potential dependence of $\sigma_{xy}$ at 0K. (c) Chemical potential dependence of $\alpha_{xy}/T$. 
  Solid, dotted, and dash-dotted line are the results for $T=$30, 50, and 100 K, respectively.
  The denoted numbers and superscripts identify the peaks and maximum or minimum in $\alpha_{xy}/T$ at which the Mott relation is violated, respectively.
  Each value of energies is tabulated in Tab.~\ref{tab:nldos_co2mnga}.
  (d) Temperature dependence of $\alpha_{xy}/T$. 
  Lines with open triangle, open circle, open square, solid triangle, and solid circle correspond to the results for $\mu=-151$, $-79$, $-62$, $-6$ and $+20$ meV, respectively.}
  \label{fig:Co2MnGa-1}
\end{figure}

\begin{figure}[tbp] \centering
  \includegraphics[width=\columnwidth]{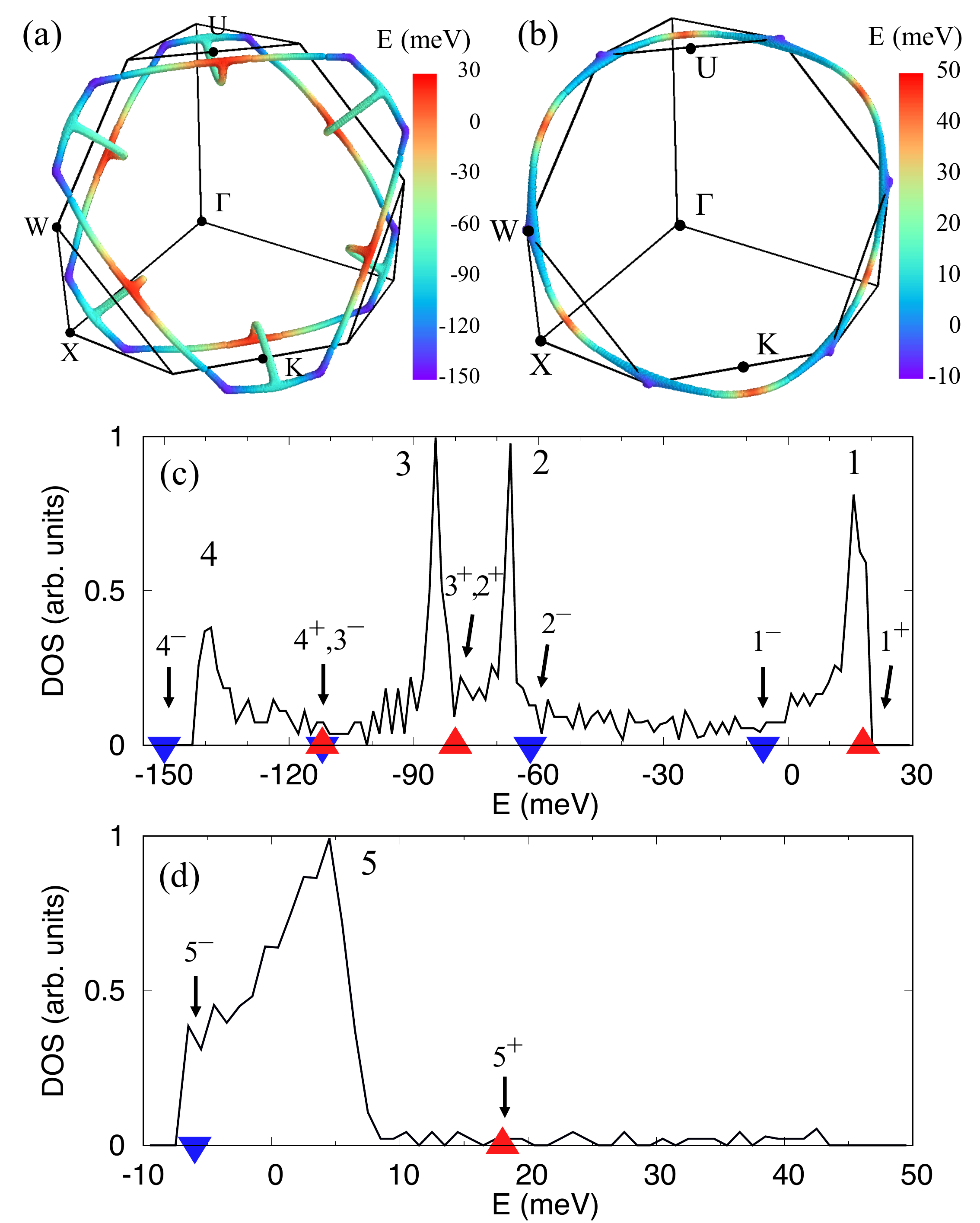}   
\caption{(a)(b) Nodal line network of Co$_2$MnGa formed by band 1 and 2, and 2 and 3 in Fig.~\ref{fig:Co2MnGa-1}(a). The color bar corresponds to the energy range from $-150$ to 50 meV, and -10 to 50 meV. (c)(d)  $D_{\rm NL}$ for the nodal line shown in (a)(b). The peak  1, 2, 3, 4 and 5 denotes $E_{\rm VHS}$ in $D_{\rm NL}$. 
  Solid upper (lower) triangle shows the maximum (minimum) in $\alpha_{xy}/T$ denoted in Fig.~\ref{fig:Co2MnGa-1}(c).}
\label{fig:Co2MnGa-NL1}
\end{figure}

Next, let us investigate the case of another prototypical magnetic Weyl semimetal Co$_2$MnGa, for which a large ANE has been recently discovered~\cite{NatPhys_Co2MnGa}.
Figure \ref{fig:Co2MnGa-1} shows the band structure, $\mu$ dependence of $\sigma_{xy}$ and $\alpha_{xy}/T$, and $T$ dependence of $\alpha_{xy}/T$.
Here we assume that the direction of the magnetization is parallel to the [001] axis.
The total magnetic moment is estimated to be $ 4.2$ $\mu_{\rm B}$/f.u., while the local magnetic moment of Mn and Co are $ 2.9$ and $0.6$ $\mu_{\rm B}$/atom, respectively.
The calculated value of $\sigma_{xy}$ at $E_{\rm F}$ (1609 $\Omega^{-1}$cm$^{-1}$) and the $\mu$ dependence of $\sigma_{xy}$ (Fig. \ref{fig:Co2MnGa-1}(b)) are consistent with the previous study.~\cite{NatPhys_Co2MnGa}
We see in Figs.~\ref{fig:Co2MnGa-1}(c) and (d) that there are several characteristic energies (20, $-6$, $-62$, $-79$ and $-151$ meV) for $\mu$ at which the absolute value of $\alpha_{xy}$ is significantly enhanced and the $\alpha_{xy}/T$ show a peculiar $T$ dependence (i.e., the Mott relation is violated). 
As we will see below, these characteristic energies can be understood in terms of the peaks in $D_{\rm NL}$.
Note that since there are five peaks in $D_{\rm NL}$ within a narrow energy range,
some of maximum and minimum points in $\alpha_{xy}/T$ are degenerated.

As has been pointed out by previous studies~\cite{NatPhys_Co2MnGa,Co2MnGa_Hasan_PRL,Guin2019}, in the low energy band structure of Co$_2$MnGa, there are several topological objects such as the Hopf link of nodal lines and type-II Weyl points. On the other hand, we are interested in the stationary points in the dispersion of the nodal lines and the relation between their energies ($E_{\rm VHS}$'s) and the characteristic energies for $\alpha_{xy}/T$. Among many nodal lines, we look into the crossing between the same spin bands. It should be noted that the effect of SOC on the crossing between the opposite spin bands is usually weak when the exchange splitting is sufficiently large. Thus the Berry curvature is expected to be large along the nodal lines made from the parallel spins. In the following, we examine the two nodal lines formed by the three bands indicated in Fig.~\ref{fig:Co2MnGa-1}(a).

In Figs.~\ref{fig:Co2MnGa-NL1}(a) and (c), we show the nodal line formed by band 1 and 2 in Fig.~\ref{fig:Co2MnGa-1}(a). From the plot of $D_{\rm NL}$ in  Fig.~\ref{fig:Co2MnGa-NL1}(c), we see that the ``band width'' of this nodal line is about 180 meV, and there are four ``van Hove singularities'' (indicated as peak 1, 2, 3, and 4) at 17, $-58$, $-85$, and $-140$ meV. 
We see that $E_{\rm VHS}$ is located between open-circle and solid-triangle points for which the Mott relation is violated.
Interestingly, there is a clear one-to-one correspondence between these $E_{\rm VHS}$'s and $E_{\rm IP}$'s (see Table \ref{tab:nldos_co2mnga})\footnote{Note that $D_{\rm NL}$ is calculated without considering SOC, so that the peaks in $D_{\rm NL}$ do not coincide perfectly with $E_{\rm IP}$ due to the SOC gap.}.
This result indicates that divergence in $D_{\rm NL}$ indeed characterizes the anomalous behavior of $\alpha_{xy}$.

It should be noted that there are type-II Weyl points whose energies are close to peak 1. In the previous study based on a model Hamiltonian~\cite{NatPhys_Co2MnGa}, it has been shown that $\alpha_{xy}/T$ shows a logarithmic divergence when the band dispersion around the Weyl points is flat and close to the transition between type-I and type-II. This result is consistent with our present argument based on the divergence in $D_{\rm NL}$.

Next, let us look into the nodal line formed by band 2 and 3 shown in Figs. \ref{fig:Co2MnGa-NL1}(b) and (d). The ``band width'' of this nodal line is just 60 meV, which is about one-third of that of the nodal line shown in Fig. \ref{fig:Co2MnGa-NL1}(a). We see that there is a peak in $D_{\rm NL}$ (peak 5 in Fig. \ref{fig:Co2MnGa-NL1}(d)) around 0 meV. This peak corresponds to the anomaly in $\alpha_{xy}/T$ at $\mu=-6$ and $18$ meV (see Table \ref{tab:nldos_co2mnga}).
While this nodal line has not been discussed in the previous study~\cite{NatPhys_Co2MnGa}, our present result suggests that hole-doping could be used to realize a large ANE in Co$_2$MnGa.

\begin{table}[tbp] \centering
  \caption{One-to-one correspondence between the peaks in $D_{\rm NL}$ and $E_{\rm IP}$ in Co$_2$MnGa.  
  $E_{\rm IP}$ is estimated as an average of energy taking the maximum and minimum in  $\alpha_{xy}/T$. 
 $\alpha_{xy}^{+(-)}/T$ denotes the energy at which $\alpha_{xy}/T$ takes its maximum (minimum) and deviates from the Mott relation.}
  \begin{tabular}{ccccc} \hline \hline
  Peak  & $D_{\rm NL}$ (meV) & $E_{\rm IP}$ (meV) & $\alpha_{xy}^+/T$ (meV) & $\alpha_{xy}^-/T$ (meV) \\ \hline 
    1    &             17    &                6      &                 -6 &   18   \\
    2    &            -68    &              -71      &                -80 &  -62   \\
    3    &            -85    &              -96      &               -112 &  -80   \\
    4    &           -140    &              -131     &               -150 & -112   \\
    5    &              4    &                6      &                 -6 &   18   \\ \hline \hline
  \end{tabular}
  \label{tab:nldos_co2mnga}
  \end{table}

\subsection{Ferromagnetic Heusler compound Fe$_3$Al}

\begin{figure}[tbp] \centering
  \includegraphics[width=\columnwidth]{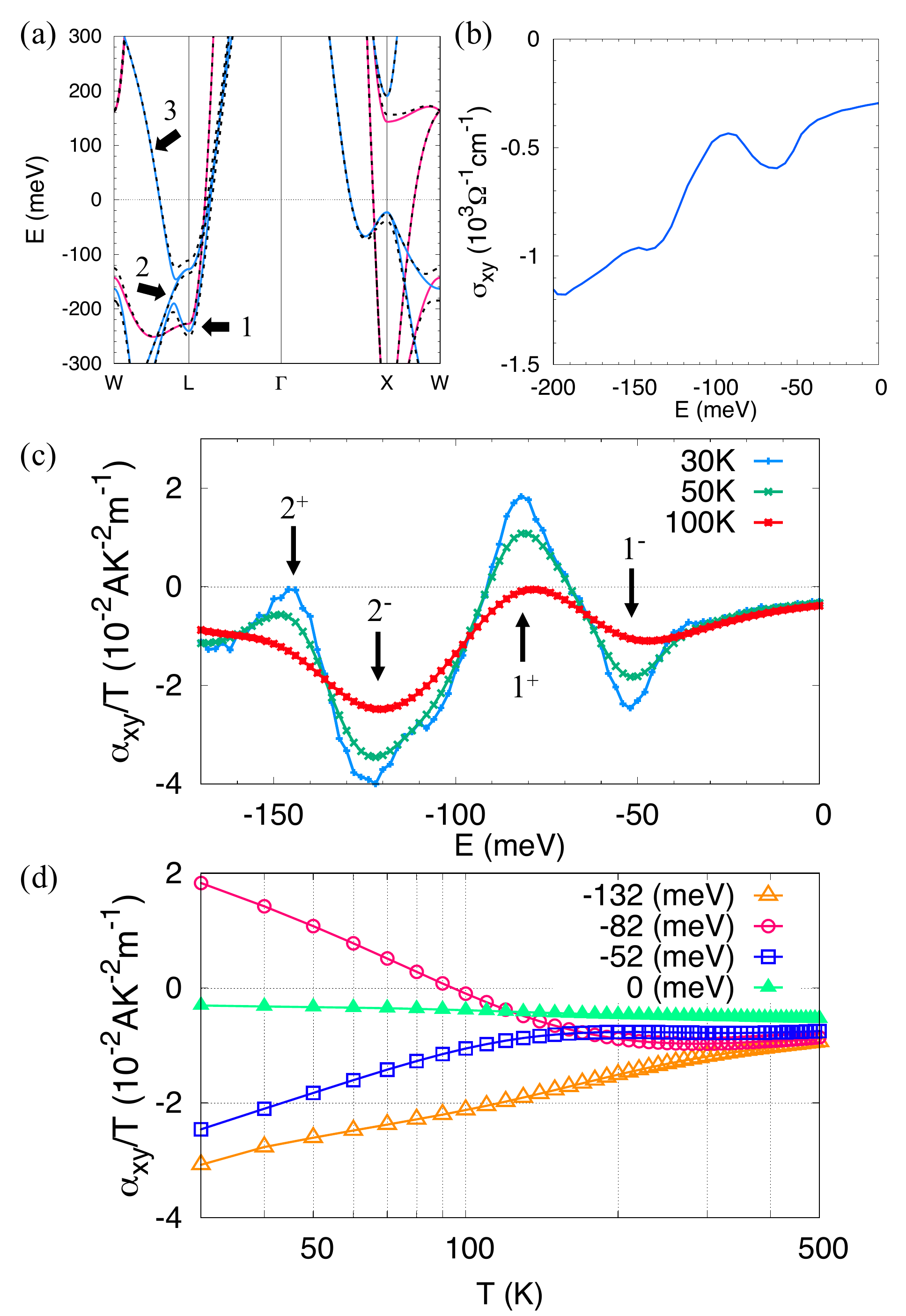}   
  \caption{(a) Band structure of \FAL. Pink and cyan lines correspond to the majority and minority spin bands computed without SOC. Dashed lines show the bands calculated including SOC. The denoted minority bands 1, 2, and 3 make the nodal lines.
  (b) Energy dependence of $\sigma_{xy}$ at 0K. (c) $\mu$ dependence of $\alpha_{xy}/T$.
  Solid, dotted, and dash-dotted line are the results for $T=$30, 50, and 100K, respectively.
  The denoted number and superscripts identify the peaks and maximum or minimum in $\alpha_{xy}/T$ at which the Mott relation is violated, respectively.
  Each value of energies is tabulated in Tab.~\ref{tab:nldos_fe3al-2}.
  (d) $T$ dependence of $\alpha_{xy}/T$. 
  Lines with open triangle, open circle, open square, and solid triangle correspond to the results for $\mu=-132$, $-82$, $-52$, and $+0$ meV, respectively.}
  \label{fig:Fe3Al-1}
\end{figure}

\begin{figure}[tbp] \centering
  \includegraphics[width=\columnwidth]{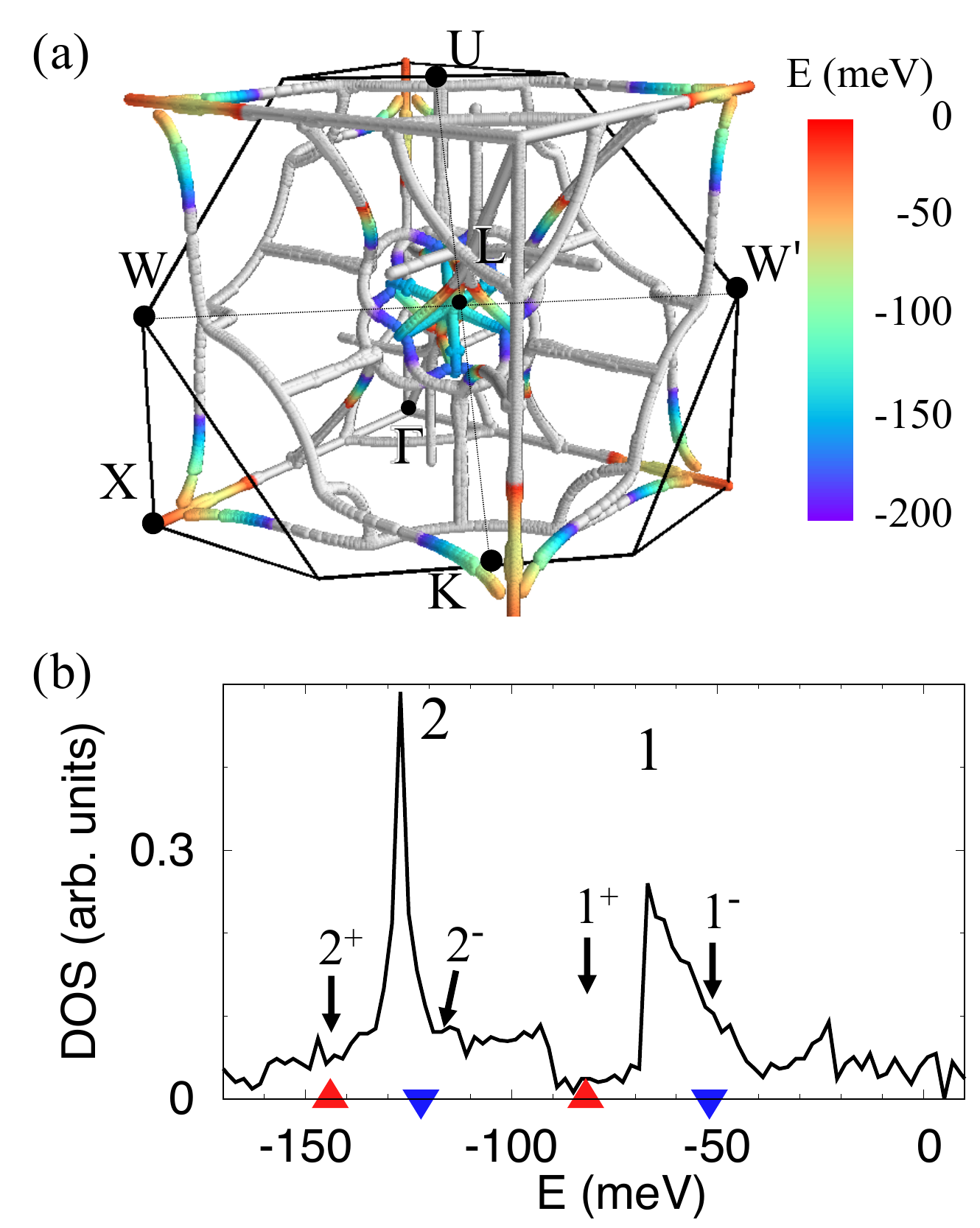}   
  \caption{
  (a) Nodal line network of Fe$_3$Al. The color bar corresponds to the energy range from $-200$ to $0$ meV. The gray part in the main figure denotes that the energy is the outside of the range of the color bar. (b) Density of states projected onto the nodal lines.
  The denoted peaks 1, 2, and 3 show the $E_{\rm VHS}$ in the $D_{\rm NL}$.
  Solid upper (lower) triangle shows the maximum (minimum) in $\alpha_{xy}/T$ denoted in Fig.~\ref{fig:Fe3Al-1}(c).}
\label{fig:Fe3Al-2}
\end{figure}

Let us finally move on to the D0$_3$-type Heusler compound Fe$_3$Al. Recently, a large ANE has been discovered in Fe$_3X$ ($X$=Al, Ga) for which the nodal lines around L point has been shown to play a crucial role\cite{Fe3X_Nature}. In the following, we focus on Fe$_3$Al and show that the enhancement of the transverse TE conductivity and the violation of the Mott relation can be understood in terms of $E_{\rm VHS}$.

Figure \ref{fig:Fe3Al-1} shows the band structure, $\mu$ dependence of $\sigma_{xy}$ and $\alpha_{xy}/T$, and $T$ dependence of $\alpha_{xy}/T$. Here we assume that the direction of the magnetization is along the [001] axis.
Regarding the atomic positions of Fe atoms, there are two types of sites Fe(I) and Fe(I\hspace{-.1em}I). For the former, Fe atoms are surrounded by other eight Fe atoms forming a cube. For the latter, Fe atoms are surrounded by other four Fe atoms and four Al atoms forming a tetrahedron.
The obtained total magnetic moment is $ 5.93$ $\mu_{\rm B}$/f.u.; the local magnetic moments of Fe(I),  Fe(I\hspace{-.1em}I), and Al are $ 2.5$, $1.9$, and $-0.3$ $\mu_{\rm B}$/atom, respectively.
These results agree well with the previous experimental and theoretical results. \cite{PhysRevB.19.2933,XU2007312}

While $\sigma_{xy}$ is just $-285$ $\Omega^{-1}$cm$^{-1}$ and its absolute value is much smaller than those of Co$_2$MnGa and Co$_3$Sn$_2$S$_2$, 
$\sigma_{xy}$ reaches $\sim -1000$ 
$\Omega^{-1}$cm$^{-1}$ when $\mu\sim -150$ meV (see 
Fig.\ref{fig:Fe3Al-1}(b)). Since $\mu$ dependence of $\sigma_{xy}$ is so drastic, we expect that the absolute value of $\alpha_{xy}$ is large.
Indeed, Figs. \ref{fig:Fe3Al-1}(c) and (d) show that while $\alpha_{xy}/T$ does not sensitively depend on $T$ for $\mu\sim 0$, $\alpha_{xy}/T$ is dramatically enhanced and the Mott relation breaks down for $\mu=-52, -82$, $-132$ and $-144$ meV.

If we look at the low-energy band structure in Fig.~\ref{fig:Fe3Al-1}(a), we see that there are many band crossings.
Among them, as in the case of Co$_2$MnGa, let us first focus on the nodal lines formed by the same spin bands. More specifically, we focus on the band 1, 2 and 3 and nodal lines formed by these bands.

Figures \ref{fig:Fe3Al-2}(a) and (b) show the nodal lines and $D_{\rm NL}$.
The nodal lines have a complex structure, which are mainly located near the high-symmetry lines such as $\Gamma$-X and $\Gamma$-L line.
There are two peaks at $-68$ and $-129$ meV in $D_{\rm NL}$ (peak 1 and peak 2).
Peak 1 (2) originates from the nodal line around the X (L) point. 
Especially, the Berry curvature is large around the L point, which is consistent with the previous results in Ref.~[\onlinecite{Fe3X_Nature}].

In Table \ref{tab:nldos_fe3al-2}, we compare $E_{\rm VHS}$ in $D_{\rm NL}$ and $E_{\rm IP}$ estimated by the energies at which $\alpha_{xy}/T$ shows a significant enhancement and the Mott relation is violated.
We see that there is a clear one-to-one correspondence between $E_{\rm VHS}$ and $E_{\rm IP}$.
This result indicates again that the divergence in $D_{\rm NL}$ generally provides useful information to search for the energy at which $\alpha_{xy}/T$ enhances dramatically.

\begin{table}[tb] \centering
 \caption{
One-to-one correspondence between the peaks in $D_{\rm NL}$ and $E_{\rm IP}$ in Fe$_3$Al. 
$E_{\rm IP}$ is estimated as an average of energy taking the maximum and minimum in  $\alpha_{xy}/T$. 
 $\alpha_{xy}^{+(-)}/T$ denotes the energy at which $\alpha_{xy}/T$ takes its maximum (minimum) and deviates from the Mott relation.}
  \begin{tabular}{ccccc} \hline \hline
    Peak  & $D_{\rm NL}$ (meV) & $E_{\rm IP}$ (meV)   & $\alpha_{xy}^+/T$ (meV)   & $\alpha_{xy}^-/T$ (meV)  \\ \hline 
    1     &               -68  &                   -67 &           -82 & -52  \\
    2     &              -129  &                  -133 &          -144 & -122 \\ \hline \hline
  \end{tabular}
  \label{tab:nldos_fe3al-2}
  \end{table}

\section{Conclusion}

In summary, to investigate the origin of the enhancement of the transverse TE conductivity ($\alpha$) in ferromagnets, we performed a systematic analysis for Co$_3$Sn$_2$S$_2$, Co$_2$MnGa and Fe$_3$Al, for which a large ANE has been recently discovered. $\alpha$ is directly related to the sum of the Berry curvature of the bands near the Fermi level. The intensity of the Berry curvature takes a large value along the nodal line, which is gapless in the non-relativistic calculation. Thus the DOS projected onto the nodal line ($D_{\rm NL}$) gives a useful information to enhance $\alpha_{ij}$. Since the nodal lines are one-dimensional objects, the stationary points in the nodal line makes a sharp ``van Hove singularities'' in $D_{\rm NL}$. When the chemical potential is close to these singularities, $\alpha_{ij}$ is dramatically enhanced. In this situation, the Mott relation breaks down and $\alpha_{ij}/T$ shows a peculiar temperature dependence. We conclude that stationary points in the nodal lines or singularities in $D_{\rm NL}$ provide a useful guide to design magnetic materials with a large ANE.

\begin{acknowledgments}
This work was supported by CREST (JPMJCR18T3, JPMJCR15Q5), by JSPS Grant-in-Aid for Scientific Research on Innovative Areas  (JP18H04481 and JP19H05825), by Grants-in-Aid for Scientific Research (JP16H06345, JP16K04875, 19H00650, 20K14390) from JSPS, and by MEXT as a social and scientific priority issue (Creation of new functional devices and high-performance materials to support next-generation industries) to be tackled by using post-K computer (hp180206 and hp190169).
The computations in this research were performed using the supercomputers at the ISSP, University of Tokyo.
\end{acknowledgments}

 \bibliographystyle{apsrev4-1}
\bibliography{ref/ref}

\end{document}